\documentclass[aps,prb,twocolumn,superscriptaddress,floatfix,longbibliography]{revtex4-2}
\usepackage{graphicx}
\usepackage{subfiles}
\usepackage{dcolumn}   
\usepackage{bm}     
\usepackage{amssymb}
\usepackage{verbatim}
\usepackage{color}
\usepackage[english]{babel}
\usepackage{hyperref}
\usepackage{amsmath}
\usepackage{amssymb}
\usepackage{derivative}
\usepackage{tikz}
\usepackage{comment}
\usepackage{wrapfig}
\usepackage{geometry}
\usepackage{subfiles}

\hypersetup{
colorlinks=true,
urlcolor= blue,
citecolor=blue,
linkcolor= blue,
}

\newif\ifptitle
\newif\ifpnumber
\newcounter{para}

\ptitletrue  
\pnumbertrue  

\begin{document}

\title{Inertia in skyrmions confined to one-dimensional geometries}

\author{Koichiro Takahashi}
\email[ ]{Koichiro.Takahashi@unh.edu}
\affiliation{Department of Physics and Astronomy, University of New Hampshire, Durham, New Hampshire 03824, USA }

\author{Sergey S. Pershoguba}
\email[ ]{pershoguba@gmail.com}
\affiliation{Department of Physics and Astronomy, University of New Hampshire, Durham, New Hampshire 03824, USA }

\author{Jiadong Zang}
\email[ ]{Jiadong.Zang@unh.edu}
\affiliation{Department of Physics and Astronomy, University of New Hampshire, Durham, New Hampshire 03824, USA }
\affiliation{Materials Science Program, University of New Hampshire, Durham, New Hampshire 03824, USA}

\date{\today}

\begin{abstract}
Magnetic skyrmions are conventionally attributed to having zero mass. In contrast, we show that skyrmions confined to one-dimensional geometries generically acquire mass (inertia) due to the combined effects of the skyrmion Hall effect and the elasticity of the system. We investigate the massive behavior of the skyrmion for a simplified periodic model of the disorder. We show that skyrmion mass lowers the critical depinning force and leads to a step-like behavior in the skyrmion velocity-vs-current curves, which were recently observed in experiments. Finite mass could also lead to hysteresis in the velocity-vs-current curves.
\end{abstract}
\maketitle

\section{Introduction}

Magnetic skyrmions~\cite{Skyrme} are prominent non-colinear spin configurations~\cite{BogdanovYablonskii, BogdanovHubert, RößlerBogdanovPfleiderer, Back_review, GOBEL20211, JIANG20171}, which are observed in a variety of noncentrosymmetric materials~\cite{Tokura_skyrmionmaterials}, e.g. $\mathrm{Mn}\mathrm{Si}$~\cite{Mühlbauer, Ishikawa_MnSi, Ishikawa_MnSi2, Lebech1995, Pfleiderer_MnSinonFermi, Grigoriev_MnFeSi, Adams_MnSi, HaifengDu_MnSinanowire, XiuzhenYu_MnSinanowire, Tonomura_MnSi, Nakajima_MnSi, Janoschek_MnSi}, $\mathrm{Fe}_{1-x}\mathrm{Co}_{x} \mathrm{Si}$~\cite{YuReal-space, Beille_FeCoSi, Ishimoto_FeCoSi, Grigoriev_FeCoSi, Grigoriev_FeCoSi2, Onose_FeCoSi, Münzer_FeCoSi, Bauer_FeCoSi, Bannenberg_FeCoSi, Morikawa_FeCoSi, Pfleiderer_2010}, $\mathrm{Fe}\mathrm{Ge}$~\cite{Lebech_1989, Uchida_FeGe, Yu2011, Wilhelm_FeGe, Gallagher_FeGe, Zhao_FeGe, Venuti_FeGe, Budhathoki_FeGe}, and $\mathrm{Mn}_{1-x}\mathrm{Fe}_x \mathrm{Ge}$~\cite{Shibata_MnFeGe, Gayles_MnFeGe, Koretsune_MnFeGe, Yokouchi_MnFeGe}. They can be stabilized at finite temperatures and driven by extremely low current densities~\cite{NagaosaTokura, Jonietz, Schulz, WeiweiWang_CurrentDynamics}. Due to their high mobility and stability, skyrmions have been a promising platform for next-generation magnetic memory devices~\cite{Parkin_memory, Fert, ZhangRacetrack, Luo_skyrmionmemory, GuoqiangYu_skyrmionmemory, Tomasello_skyrmionmemory, Koshibae_skyrmionmemory, KyungMeeSong_skyrmionmemory, Wang_skyrmionmemory, Sampaio_memory, Romming_memory, Lai_edgeconfinement, Fook_racetrack}.  

The dynamics of skyrmions are governed by the Landau-Lifshitz-Gilbert (LLG) equation. A conventional way to simplify this equation is by using Thiele's method~\cite{Thiele, Everschor_Thiele}. This approach treats a skyrmion as a rigid particle~\cite{Shi-ZengLin_Thiele} described by a two-dimension position vector $\bm r(t)$. As a result, the LLG equation reduces to a first-order differential equation in terms of $\dot {\bm r}(t)$. The reduced equation allows us to describe the current-induced motion of the skyrmion~\cite{Thiaville_ThieleModified, IwasakiCurrent, IwasakiUniversal}. Experimental and theoretical studies have shown that, in response to an applied electric current density $\bm j$, skyrmions acquire velocities with components both parallel and perpendicular to $\bm j$. The latter effect is known as the skyrmion Hall effect~\cite{Zang_skyrmionHall, Jiang_skyrmionHall, Litzius_skyrmionHall, Juge_skyrmionHall, Woo_skyrmionHall}.
\begin{figure}[hbt!]
    \centering
    (a) \includegraphics[width=0.8\linewidth]{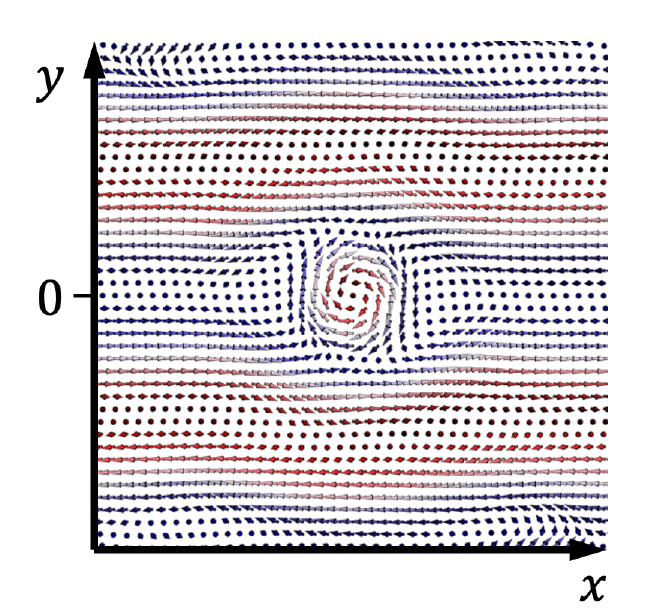} \\
    (b) \includegraphics[width=0.8\linewidth]{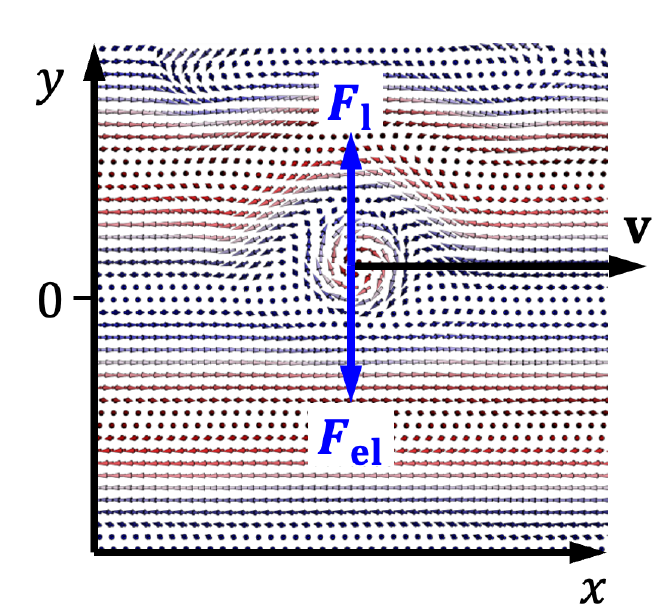}
     \caption{Graphical representation of the system. (a) A skyrmion confined inside a helical lane at rest. (b) A finite current density $\bm j \propto \hat{\bm x}$ pushes the skyrmion with finite velocity $\bm v \propto \hat{\bm x}$ along the lanes. The skyrmion shifts to a new equilibrium vertical position $y$, where the Lorentz force $F_\mathrm{L} \propto v$ is balanced out by the elastic force $F_{\mathrm{el}} \propto y$ produced by the deformed lane. Since $F_{\mathrm{el}} = F_{\mathrm{L}}$, the elastic potential energy scales as $U_{\mathrm{el}} \propto y^2 \propto v^2$, which can be interpreted as a kinetic energy giving rise to the skyrmion inertia.
     }  
     \label{fig:skyrmion}
\end{figure}

The fact that the Thiele equation is first-order in the derivative $\dot {\bm r}(t)$ is usually interpreted as the absence of skyrmion mass, i.e. $m_{\mathrm{sk}}=0$. However, it has been shown that skyrmion mass can effectively be generated~\cite{Schütte, Makhfudz_Inertia, XiaofanWu_massivemassless, Shi-ZengLin_Inertia} due to the deformation of its shape. Skyrmion mass could have nontrivial implications for their current-induced dynamics. The purpose of this work is to point out that skyrmion mass is generically generated when skyrmions are confined to quasi-one-dimensional geometries (1D). The mass arises due to the combined effects of the skyrmion Hall effect and the system's elasticity. To date, there are two established approaches to confining skyrmions to 1D: using helical backgrounds~\cite{Ezawa_helicalbackground, Müller_helical, Song2022, LingyaoKong_Helical, Schoenherr_nanostripe, Weienhofer_helical, Schick_helical, GuangYao_antiskyrmionHelical} or using magnetic nanostripes~\cite{HaifengDu_nanostripe, ZhangRacetrack, Knapman_Nanostripe, Song2024, Purnama_1Dwall, Müller_two-lane, Lai_edgeconfinement, DongLiang_nanowire}. In this work, we focus on the former scenario, however the results are equally applicable to the latter.  

Consider a skyrmion nucleated inside the helical lanes~\cite{Müller_helical, Song2022, LingyaoKong_Helical, Schoenherr_nanostripe, Weienhofer_helical, Schick_helical} illustrated in Fig.~\ref{fig:skyrmion}. At rest, the skyrmion occupies a symmetric position ($y=0$) between the two elastic helical lanes as shown in Fig.~\ref{fig:skyrmion}(a). At finite horizontal velocity $v$, as illustrated in Fig.~\ref{fig:skyrmion}(b), the skyrmion experiences the Lorentz force $F_{\rm{L}}$ due to the skyrmion Hall effect, which pushes the skyrmion along the positive $y$-axis. The helical lane acts as an elastic rubber band obstructing the skyrmion motion along $y$-axis. In response to the pressure from the skyrmion, the top lane deforms and pushes back with the elastic Hooke's force $F_{\rm{el}}$. The skyrmion finds a new equilibrium position $y\neq 0$, where the Lorentz and elastic force balance each other $F_{\rm{L}} = F_{\rm{el}}$. Since the former force is proportional to the horizontal velocity $F_{\rm{L}}\propto v$, whereas the latter force is proportional to the vertical displacement $F_{\rm{el}} \propto y$, we conclude that $y\propto v$. So, the potential energy of elastic deformation is proportional to the square of velocity $U_{\rm{el}} \propto y^2 \propto v^2$ and, therefore, can be interpreted as a kinetic energy leading to skyrmion inertia. The same phenomenology applies to skyrmions constricted within narrow nanoribbons~\cite{HaifengDu_nanostripe, Knapman_Nanostripe, Song2024, Purnama_1Dwall, Müller_two-lane, DongLiang_nanowire}, as long as the skyrmions are repelled from the nanoribbon's edge but not annihilated by it~\cite{IwasakiCurrent}. Real materials include impurities and disorders, which will cause skyrmion pinning at low current densities~\cite{Hoshino_disorder, Reichhardt_disorderreview}. For this reason, we include a simplified disorder potential in our model. We note that a similar mass generation mechanism was discussed in 2D nanodisc geometries~\cite{XiaofanWu_massivemassless, Ivanov_Inertia, Kyoung-WoongMoon_Inertia, Myoung-WooYoo_Inertia, Shiino_Inertia}.

Below, we analyze the current-driven dynamics of a skyrmion confined to a quasi-1D lane by using the Thiele equation derived in Sec.~\ref{sec:Thiele}. In Sec.~\ref{sec:deriving_mass}, we derive the non-zero skyrmion mass $m_{\rm sk}$. In Sec.~\ref{sec:periodic_disorder}, we apply our formalism to a simplified periodic model of the disorder. We find that skyrmion dynamics can be classified by a parameter $\tilde m_{\rm sk}$ representing a dimensionless skyrmion mass, given by Eq.~(\ref{lambda}). We identify two regimes for the skyrmion dynamics: of the (overdamped) small mass $\tilde m_{\rm sk}\ll 1$  and the (underdamped) large mass $\tilde m_{\rm sk}\gg 1$ regimes. In the latter regime, we show that the skyrmion dynamics could exhibit hysteresis in $v_{\rm sk}$-vs-$j$ graph akin to a similar behavior in Josephson junctions~\cite{Tinkham}.
\section{Equation of skyrmion motion: Thiele equation} \label{sec:Thiele}

To analyze the dynamics of the magnetization $\bm M(\bm r,t)$ driven by the electric current density $\bm j$, the standard classical starting point is the Landau-Gilbert-Lifshitz (LLG) equation
\begin{align}
\frac{\partial \bm{M}}{\partial t} = & \,\gamma  \bm B^{\mathrm{eff}}\times \bm M + \frac{\alpha}{M} \bm{M} \times \frac{\partial \bm{M}}{\partial t} \label{LLG_main} \\
& + \frac{p a^3}{2 e M} \left( \bm{j} \cdot \nabla \right) \bm{M} - \frac{p a^3 \beta}{2 e M^2} \left[ \bm{M} \times \left( \bm{j} \cdot \nabla \right) \bm{M} \right]. \nonumber
\end{align}
For simplicity, we assume a planar geometry with $\bm r = (x,y)$ representing two-dimensional (2D) coordinates and $\bm M = (M_x,M_y,M_z)$ is a three-dimensional (3D) magnetization vector. The first term ($\propto \gamma$) on the right-hand side describes the gyromagnetic coupling with local magnetic field $\bm{B}^{\mathrm{eff}}(\bm{r},t)$, which can be derived from the spin Hamiltonian of the system~\cite{IwasakiCurrent}. The second term ($\propto\alpha$) describes the Gilbert damping~\cite{Gilbert_damping} and produces friction. The third and fourth terms in Eq.~(\ref{LLG_main}) are known as the spin-transfer torques (STT)~\cite{Slonczewski}. They describe the coupling with electric current density $\bm j$. 

While the partial differential equation~(\ref{LLG_main}) is amenable for numerical integration, an adequate approximate solution can be obtained by Thiele's approach~\cite{Thiele}. In this approach, it is assumed that the skyrmion can be treated as a rigid particle. So the magnetization dynamics may be approximated by an ansatz
\begin{equation}
    \bm M(\bm r',t) = \bm M[\bm r'-\bm r(t)], \label{mrt}
\end{equation}
where $\bm r(t) = [x(t),y(t)]$ represents the time-dependent 2D position of the skyrmion. Substituting Eq.~(\ref{mrt}) into Eq.~(\ref{LLG_main}) and it reduces to a first-order differential equation 
\begin{align}
\mathcal G\,\hat {\bm z} \times \left( \bm u - \dot{\bm r} \right) + \mathcal{D} \left( \beta \, \bm u - \alpha  \, \dot{\bm r} \right) - \bm\nabla U(\bm r) =\bm{0} \label{Thiele_basic}
\end{align}
in $\bm{r}(t)$. The constants $\alpha$ and $\beta$ carry over from Eq.~(\ref{LLG_main}). The vector $\bm u = - \frac{p a^3}{2 e M} \bm j$ is conventionally interpreted as the velocity of spins conducted by itinerant electrons. The two dimensionless constants 
\begin{align*}
& \mathcal G   = \int d^2 r \,\, \hat{\bm \Omega} \cdot \left(\pdv{\hat{\bm \Omega}}{x} \times {\pdv{\hat{\bm \Omega}}{y}} \right) = 4\pi, \quad \hat{\bm \Omega} = \frac{\bm M}{M}, \\
&\mathcal{D} = \frac{1}{2}\int d^2 r  \left[\left(\pdv{\hat{\bm{\Omega}}}{x} \right)^2 + \left(\pdv{\hat{\bm{\Omega}}}{y} \right)^2\right],  
\end{align*}
describe the geometry of the skyrmion configuration. The constant $\mathcal G$ represents the topological number of the skyrmion, which is independent of small deformations of the skyrmion shape caused by its motion. In contrast, constant $\mathcal D$ may depend on the skyrmion shape, but that effect is not essential for the discussion below. The phenomenological potential energy $U(\bm r)$ is added to account for the effects of the disorder and the confinement by helical lanes. Note that $U(\bm r)$ has unconventional units of length$^2$/time.

A notable feature is that Eq.~(\ref{Thiele_basic}) contains the first-order $\dot {\bm r}$ but not the second-order $\ddot {\bm r}$ derivative. This fact is usually interpreted in literature as the absence of skyrmion mass. Nevertheless, as shown below, the skyrmion mass $m_{\rm sk}$ is effectively generated due to the combined effects of the skyrmion Hall effect and the elasticity of helical lanes.

\section{Skyrmion mass} \label{sec:deriving_mass}

The potential $U(\bm r)$ appearing in Eq.~(\ref{Thiele_basic}) accounts both for helical lanes, illustrated in Fig.~\ref{fig:skyrmion}, and the disorder experienced by the skyrmion. To proceed further, we make a simplifying assumption that the potential is separable 
\begin{align}
U(\bm r) =  U_\mathrm{hel}(y) + U_\mathrm{dis}(x) \label{total_potential}
\end{align}
in $x$ and $y$ variables. The first term $U_\mathrm{hel}(y)$ describes the confinement potential due to helical lanes. At small $y$, the potential is parabolic
\begin{align}
U_\mathrm{hel}(y) = \frac{1}{2} k \,y^2 \label{helical_potential}
\end{align}
where $k$ is the corresponding elasticity constant. Assuming that the helical potential~(\ref{helical_potential}) is sufficiently stiff, it confines the skyrmion within one helical lane $|y|<w$, where $w$ is the lane width. Consequently, the skyrmion is subject to the disorder only within the corresponding helical lane, and we can model the disorder potential as a purely $x$-dependent function $U_\mathrm{dis}(x)$. 

We assume that the current $\bm j$ is pointing along the helical lanes
\begin{equation}
  \bm u = u \, \hat{\bm x}. \label{current}
\end{equation}
Substituting Eqs.~(\ref{total_potential}), (\ref{helical_potential}), and (\ref{current}) into the Thiele equation (\ref{Thiele_basic}) and writing it in components, we obtain a pair of first-order differential equations
\begin{equation}
\begin{aligned}
   & \mathcal G \,\dot y + \mathcal D (\beta\, u - \alpha \,\dot x) - \frac{\partial U_\mathrm{dis}(x)}{\partial x} = 0, \\
   & \mathcal G (u-\dot x) - \mathcal D \,\alpha\, \dot y - k\, y = 0.
\end{aligned} 
\label{Thiele_two_component}
\end{equation}
Eq.~(\ref{Thiele_two_component}) can be reduced to a single second-order differential equation on $x(t)$. Differentiating Eq.~(\ref{Thiele_two_component}) over time, we obtain a system of four equations in total. Eliminating the variables $y(t)$, we arrive at a second-order differential equation on $x(t)$ (See Appendix.~\ref{sec:derivation_1d_eom})
\begin{widetext}
\begin{align}
 \left(\frac{\mathcal{G}^2 + \mathcal{D}^2\alpha^2}{k}\right) \ddot{x} + \mathcal{D} \alpha\left(1 + \frac 1k\frac{\partial^2 U_\mathrm{dis}(x)}{\partial x^2}\right)\dot{x} - \left(\frac{\mathcal{G}^2 + \mathcal{D}^2\alpha\beta}{k}\right)\dot{u} - \mathcal{D} \beta u + \frac{\partial U_\mathrm{dis}(x)}{\partial x} = 0. \label{1d_eom}
\end{align}
\end{widetext}
Let us comment on the terms appearing in Eq.~(\ref{1d_eom}), which is the main focus of this paper. The equation has a form of Newton's 2nd law with the combination 
\begin{align}
m_\mathrm{sk} = \frac{\mathcal{G}^2 + \mathcal{D}^2\alpha^2}{k} \label{sk_mass}
\end{align}
playing the role of skyrmion mass. In realistic experimental situations, the Gilbert damping is small $\alpha \ll 1$, so the skyrmion mass $m_\mathrm{sk} \approx \mathcal G^2/k$ is determined by the skyrmion Hall effect $\mathcal G$ and the elasticity constant $k$, and the mass has units of time. The last three terms in Eq.~(\ref{1d_eom}) represent the three sources of driving force acting on the skyrmion. The term $\partial U_\mathrm{dis}(x)/\partial x$ describes the effect of the disorder. The two terms $\propto u$ and $\propto \dot u$ represent the force produced by the electric current $u(t)$, which we henceforth refer to as the ``push force''. The second term $\propto \dot x$ describes a variable friction force experienced by the interstitial skyrmion.  

Let us comment on the applicability of our approach. Our model relies on quadratic approximation~(\ref{helical_potential}) for the potential induced by helical lanes. So, the model is not applicable, when that approximation fails, i.e. where the vertical displacement $y$ is of the order of interlane distance $w$, i.e. $y \sim w$. This will occur at some maximum push force $u_\mathrm{max}$ causing the interstitial skyrmion to be pressed too hard against the helical lanes due to the skyrmion Hall effect. The value of $u_{\rm max}$ can be estimated by dropping the disorder potential from Eq.~(\ref{Thiele_two_component}), setting $y=w$, $\dot y = 0$, ignoring the disorder term due to the large current, and solving for $u$
\begin{equation}
u_\mathrm{max} \sim \frac{\alpha}{\alpha - \beta} \frac{k w}{\mathcal{G}}.
\label{upper_lim}
\end{equation}
We expect that our approach is applicable at push force below $u<u_\mathrm{max}$. In practice, experiments show that large current densities $j$ cause skyrmions to annihilate or transform into meron pairs \cite{Song2022,Song2024}. 

\section{Skyrmion dynamics in a case of a simplified disorder potential} \label{sec:periodic_disorder}

\begin{figure}
    \includegraphics[width=0.9\linewidth]{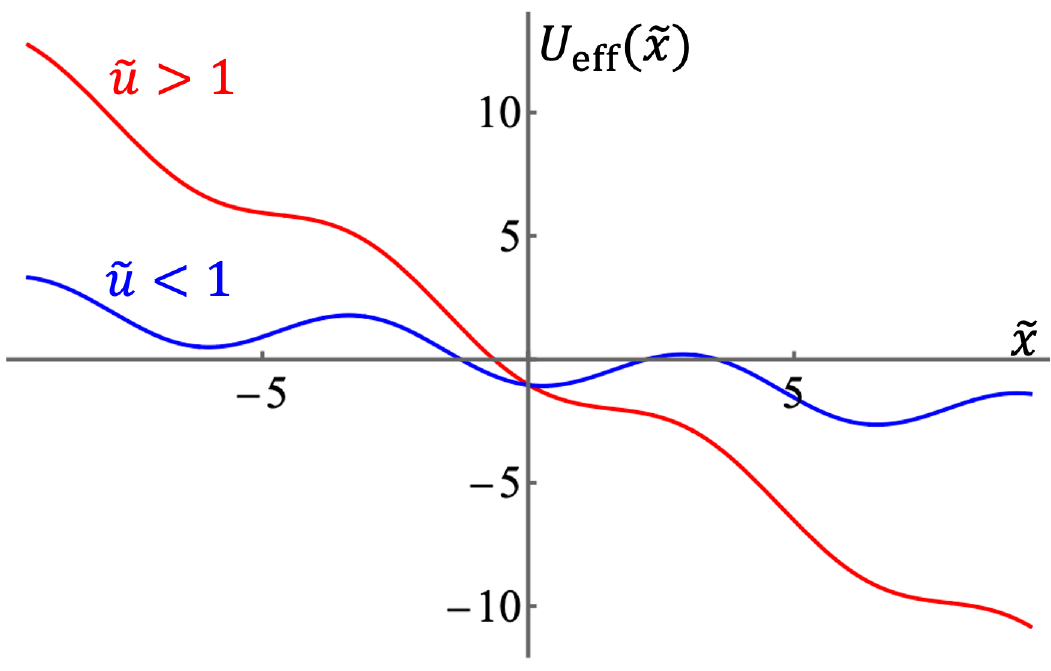}  
      \caption{If the push force is large, i.e. $\tilde u>1$, the washboard potential is monotonic causing the skyrmion to propagate. In contrast, if the push force is small, i.e. $\tilde u<1$, the washboard potential is non-monotonic with minima, which could pin the skyrmion.} 
      \label{fig:washerboard}
\end{figure}

The purpose of this section is to investigate skyrmion dynamics for a simple model of the disorder, however, all conclusions carry over to more realistic models. (i) We assume that $u(t)$ is varied slowly, so $\dot u$ in Eq.~(\ref{1d_eom}) can be neglected. (ii) We assume that stiffness of helices $k$ is much greater than that of the disorder $\partial^2 U_\mathrm{dis}(x)/\partial x^2$, so their ratio $\partial^2 U_\mathrm{dis}(x)/k\partial x^2$ in the friction term can be neglected. (iii) Finally, we model the disorder potential as a periodic function~\cite{Reichhardtonedimentional, Yi-fuChen_Harmonic} 
\begin{align}
U_\mathrm{dis}(x) = - U_0 \cos{K x},
\end{align}
where $U_0$ describes the disorder strength, and $\tilde m_\mathrm{sk} = \frac{2\pi}{K}$ represents a typical spatial scale over which the disorder varies. Equivalently, the wave number $K$ characterizes the average density of impurities. Then, Eq.~(\ref{1d_eom}) reduces to  
\begin{align}
 m_{\mathrm{sk}}\,\ddot{x} + \mathcal{D} \alpha \,\dot{x} - \mathcal{D} \beta u + U_0 K \sin(Kx) = 0.
\label{1d_eom_cosine_disorder}
\end{align}
To analyze this equation, it is practical to switch to dimensionless variables denoted by ``$\sim$'':
\begin{equation}
\begin{aligned}
    \tilde x = K\,x, \quad
    \tilde t = \dfrac{t}{\mathcal D\alpha/U_0K^2}, \quad
    \tilde u = \dfrac{u}{U_0K/\mathcal D\beta}.
\end{aligned}
\end{equation}
In the new variables, Eq.~(\ref{1d_eom_cosine_disorder}) simplifies to
\begin{align}
 \tilde m_\mathrm{sk}\frac{d^2\tilde x}{d\tilde t^2} + \frac{d\tilde x}{d\tilde t} - \tilde u + \sin\tilde x = 0,
\label{1d_eom_cosine_disorder_dimensionless}
\end{align}
with a constant
\begin{equation}
\tilde m_\mathrm{sk} = \frac{m_{\mathrm{sk}}\, U_0 K^2}{\alpha^2\, \mathcal D^2} \label{lambda}
\end{equation}
representing a dimensionless mass. 

It is instructive to combine the last two terms of Eq.~(\ref{1d_eom_cosine_disorder_dimensionless}) into a single effective potential 
\begin{align}
& \tilde m_\mathrm{sk}\frac{d^2\tilde x}{d\tilde t^2} + \frac{d\tilde x}{d\tilde t} + \frac{\partial U_{\rm eff}(\tilde x)}{\partial \tilde x} = 0, \nonumber \\
& U_\mathrm{eff}(\tilde x) = - \,\tilde u \,\tilde x - \cos \tilde x. \label{washerboard}
\end{align}
It is known as a washboard potential and has been extensively studied in the context of Josephson junctions~\cite{Tinkham, Strogatz, NagaosaQFT}. The washboard potential consists of a linear term in $\tilde x$ and a periodic potential due to the disorder as shown in Fig.~\ref{fig:washerboard}. If the push force is larger than the disorder potential, i.e. $\tilde u > 1$, the potential is a monotonously decreasing function causing the skyrmion to roll down the hill. In contrast, if the push force is smaller, i.e.  $\tilde u < 1$, the potential becomes non-monotonic and develops local minima as shown by the blue curve. In the absence of skyrmion mass, i.e. $m_\mathrm{sk} = 0$, the potential minima inevitably pin the skyrmion. However, in the presence of the skyrmion mass, i.e.  $m_\mathrm{sk} \neq 0$, the skyrmion could sustain a propagating mode due to its inertia allowing the skyrmion to overcome potential barriers. 

The discussion below follows a known behavior in Josephson junctions~\cite{Tinkham}, according to which, the dynamics can be classified as either overdamped friction-dominated $\tilde m_\mathrm{sk} \ll 1$ or underdamped mass-dominated $\tilde m_\mathrm{sk} \gg 1$.

\subsection{Overdamped regime $\tilde m_\mathrm{sk} \ll 1$}

We set $\tilde m_\mathrm{sk} = 0$, so Eq.~(\ref{1d_eom_cosine_disorder}) becomes
\begin{align}
  \frac{d\tilde x}{d\tilde t} =  \tilde u - \sin\tilde x. \label{1d_eom_massless}
\end{align}
It implies that the velocity of the skyrmion is a periodic function of $\tilde x$. Integrating this first-order differential equation (see Appendix.~\ref{sec:k_inf_limit}), we compute the velocity averaged over time
\begin{align}
 \langle \tilde{v}_{\rm sk}
 \rangle \equiv \left\langle \frac{d\tilde x}{d\tilde t}\right\rangle_t = \left\{
\begin{array}{ll}
    0 &  \text{if } \tilde u \leq 1,\\[1em]
   \sqrt{ \tilde u^2 - 1} &  \text{if } \tilde u > 1.
\end{array}
\right. \label{current_velocity_large_friction}
\end{align}
At low push force $\tilde u \le 1$, the skyrmion is pinned by the minima in the washboard potential~(\ref{washerboard}). Above the critical depinning force  $\tilde{u} > 1$, the minima of the washboard potential~(\ref{washerboard}) disappear, and it becomes a monotonic function causing the skyrmion to propagate with finite velocity.


\subsection{Underdamped regime $\tilde m_\mathrm{sk} \gg 1$.} \label{sec:large_lambda}
The discussion above shows that massless skyrmion is pinned if the washboard potential contains barriers at $\tilde u<1$. Below, we show that, in the opposite limit of large mass $\tilde m_\mathrm{sk} \gg 1$, the skyrmion can overcome the barriers using its inertia. This will cause lower critical depinning currents $\tilde u_{\rm cr} < 1$. 

To proceed further, we move the second and the third term to the right-hand side of Eq.~(\ref{1d_eom_cosine_disorder_dimensionless}), multiply the whole equation by $d{\tilde x}/d\tilde t$ and integrate:
\begin{align}
   &\frac 12 \tilde m_\mathrm{sk}\, \tilde{v}_{\mathrm{sk}}^2(\tilde{x}) -  \cos \tilde x - \left(\varepsilon_0 - 1\right) = \int^{\tilde{x}}_{0} dx\,\left[ \tilde u -   \tilde{v}_{\mathrm{sk}}(x)\right].  \label{energy_eq} 
\end{align}
Here, we switched to a new independent variable $\tilde x$ instead of $\tilde t$ and denoted $d{\tilde x}/d\tilde t \equiv \tilde{v}_{\mathrm{sk}}$. Eq.~(\ref{energy_eq}) has a transparent physical meaning. It relates the change of the mechanical energy (on the left-hand side) to the work done by the push $\propto\tilde u$ and the friction  $\propto \tilde{v}_{\mathrm{sk}}(\tilde{x})$ forces (on the right-hand side). The constant of integration $\varepsilon_0$ represents the initial kinetic energy of the skyrmion. 

Eq.~(\ref{energy_eq}) can be viewed as an integral equation on the unknown function $\tilde{v}_{\mathrm{sk}}(\tilde{x})$. 
We focus on a periodic solution~\cite{Strogatz}, where the net work in the right-hand side of Eq.~(\ref{energy_eq}) is zero over one period
\begin{align}
  \int^{2\pi}_0 d\tilde x\, \left[ \tilde u -   \tilde{v}_{\mathrm{sk}}(\tilde x)\right] = 0. \label{net_work_zero}
\end{align}
Integrating the term $\propto \tilde u$  allows us to cast Eq.~(\ref{net_work_zero}) in a form
\begin{align}
  \tilde u = \frac{1}{2\pi}\int^{2\pi}_0 d\tilde x\, \tilde{v}_{\mathrm{sk}}(\tilde x) . \label{u_vs_vsk}
\end{align}
In the limit of small friction $\tilde m_\mathrm{sk} \gg 1$, on the other hand, the mechanical energy [the left-hand side of Eq.~(\ref{energy_eq})] is conserved $\frac 12 \tilde m_\mathrm{sk} \,\tilde{v}_{\mathrm{sk}}^2(\tilde x) - \cos \tilde x - \varepsilon_\ast(\tilde u) +1 = 0$ to a good accuracy at arbitrary $\tilde{x}$. Here $\varepsilon_\ast(\tilde u)$ denotes the kinetic energy of the periodic orbit that replaced the arbitrary constant $\varepsilon_0$. Solving the latter algebraic equation in terms of velocity, one finds
\begin{align}
   \tilde v_\mathrm{sk}(\tilde x) = \sqrt{\frac 2{\tilde m_\mathrm{sk}} \left( \varepsilon_\ast(\tilde u) - 2\sin^2\frac {\tilde x}2 \right)}. \label{vskx}
\end{align}
Substituting it into Eq.~(\ref{u_vs_vsk}), we obtain
\begin{align}
  \tilde u = \frac{1}{\pi}\sqrt{\frac{8\,\varepsilon_\ast(\tilde u)}{\tilde m_\mathrm{sk}}}\,E\left(\sqrt{\frac{2}{\varepsilon_\ast(\tilde u)}}\right),  \label{u_vs_e} 
\end{align}
using a complete elliptic integral of the second kind $E(z)$ \footnote{\label{elliptic}We use the following definitions of elliptic integrals $K(z) = \int_0^{\pi/2} \frac{dx}{\sqrt{1-z^2 \sin^2x}}$ and $E(z) = \int_0^{\pi/2} dx\sqrt{1-z^2 \sin^2x}$}. This implicit equation can be solved numerically to find the energy   $\varepsilon_\ast(\tilde u)$ corresponding to push force $\tilde u$. If $\varepsilon_\ast(\tilde u)$ is known, one can substitute it back into Eq.~(\ref{vskx}) and compute the velocity averaged over time
\begin{align}
\langle \tilde v_\mathrm{sk} \rangle = \left\{
\begin{array}{ll}
    0 &  \text{if } \tilde u < \tilde u_{\mathrm{cr}}\\[1em]
    \sqrt{\frac{\varepsilon_\ast(\tilde u)}{2 \tilde m_\mathrm{sk}}} \cdot \frac{\pi}{K\left(\sqrt{\frac{2}{\varepsilon_\ast(\tilde u)}}\right)} & \text{if } \tilde u > \tilde u_{\mathrm{cr}}. 
\end{array}
\right.  \label{current_velocity_large_mass}
\end{align}
where $ 
   \tilde u_{\mathrm{cr}} = \frac{4}{\pi}\,\frac 1{\sqrt{\tilde m_\mathrm{sk}}}. \label{ucr2} $
is a critical push force at which depinning occurs, and $K(z)$ is a complete elliptic integral of the first kind. A pair of equations~(\ref{u_vs_e}) and (\ref{current_velocity_large_mass}) should be used to compute the $\tilde v_\mathrm{sk}$-vs-$\tilde u$ curves numerically.

\begin{figure}
    \centering
     \includegraphics[width=\linewidth]{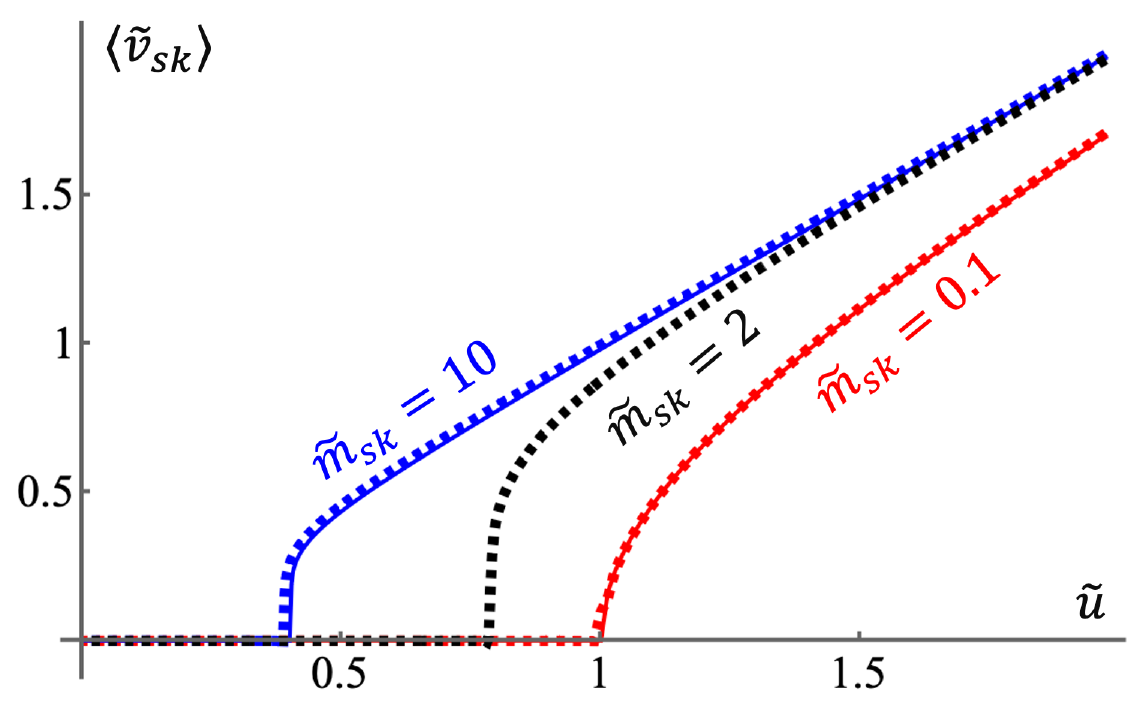}   
      \caption{The  $\tilde v_{\rm sk}$-vs-$\tilde u$ curves for a skyrmion in a quasi-1D lane. The dashed lines correspond to the numerical solutions of the differential equation~(\ref{1d_eom_cosine_disorder_dimensionless}), whereas the solid lines correspond to the analytical expressions~(\ref{current_velocity_large_friction}) and (\ref{current_velocity_large_mass})} 
      \label{fig:u_vs_v}
\end{figure}
Now, let us compute the $\tilde v_{\mathrm{sk}}$-vs-$\tilde u$ curves for a range of parameters $\tilde m_\mathrm{sk}$ corresponding to friction-dominated $\tilde m_\mathrm{sk}\ll 1$ and mass-dominated $\tilde m_\mathrm{sk}\gg 1$ regimes and present the results in Fig.~\ref{fig:u_vs_v}. We solve the differential equation~(\ref{1d_eom_cosine_disorder_dimensionless}) numerically, average the velocity over time $\langle \tilde v_{\mathrm{sk}}\rangle = \langle \dot{\tilde x}\rangle_t$ and plot it in dashed lines. For comparison, the analytical expressions~(\ref{current_velocity_large_friction}) and (\ref{current_velocity_large_mass}), plotted in solid lines, agree with the numerical results well. At large current $\tilde u$, all curves approach the same asymptote $\langle \tilde v_{\mathrm{sk}}\rangle = \tilde u$. The main effect of skyrmion inertia appears near depinning. (i) In the massless regime, $\tilde m_\mathrm{sk} = 0$, the critical depinning current is set by the disorder potential $\tilde u_{\mathrm{cr}} = 1$ (or $u_\mathrm{cr} = \frac{U_0 K}{\mathcal D \beta}$ in the dimensional units). The velocity grows as a square root $\tilde v_\mathrm{sk}(\delta \tilde u) \sim \sqrt{2\,\delta \tilde u}$ close to depinning $\delta \tilde u = \tilde u - \tilde u_\mathrm{cr} \ll 1$. (ii) In the large mass limit $\tilde m_\mathrm{sk} \gg 1$, the critical depinning current is lowered to $\tilde u_\mathrm{cr} = \frac{4}{\pi \sqrt{ \tilde m_\mathrm{sk}}}$ ($u_\mathrm{cr} = \frac{4\alpha}{\pi \beta}\,\sqrt{\frac{U_0}{m_\mathrm{sk}}}$ in the dimensional units). The velocity exhibits a strongly non-analytic step-like behavior. It can be expressed $\tilde v_\mathrm{sk}(\delta \tilde u) \sim \frac{1}{\sqrt{\tilde m_\mathrm{sk}} W_{-1}(-\sqrt{ \tilde m_\mathrm{sk} \delta \tilde u})}$, via a Lambert function $W(x)$~\footnote{The Lambert function $W(x)$ is defined as a the solution of the equation $W(x) e^{W(x)} = 1$.  We are interested in the real branch $W_{-1}(x)$, which diverges at $W_{-1}(x) \to -\infty$ at $x\to -0$.}. Such a sharp step-wise onset of the velocity near depinning may serve as an indicator of the underdamped regime $\tilde m_\mathrm{sk} \gg 1$. A similar step-like behavior was recently observed in Ref.~\cite{Song2024}, which we take as a strong indication that $\tilde m_\mathrm{sk} \gg 1$ in that system.

\subsection{Hysteresis in the $\tilde v_{\mathrm{sk}}$-vs-$\tilde u$ curves at $\tilde m_\mathrm{sk} \gg 1$}
\begin{figure}
    \centering
  (a)  \includegraphics[width=0.9\linewidth]{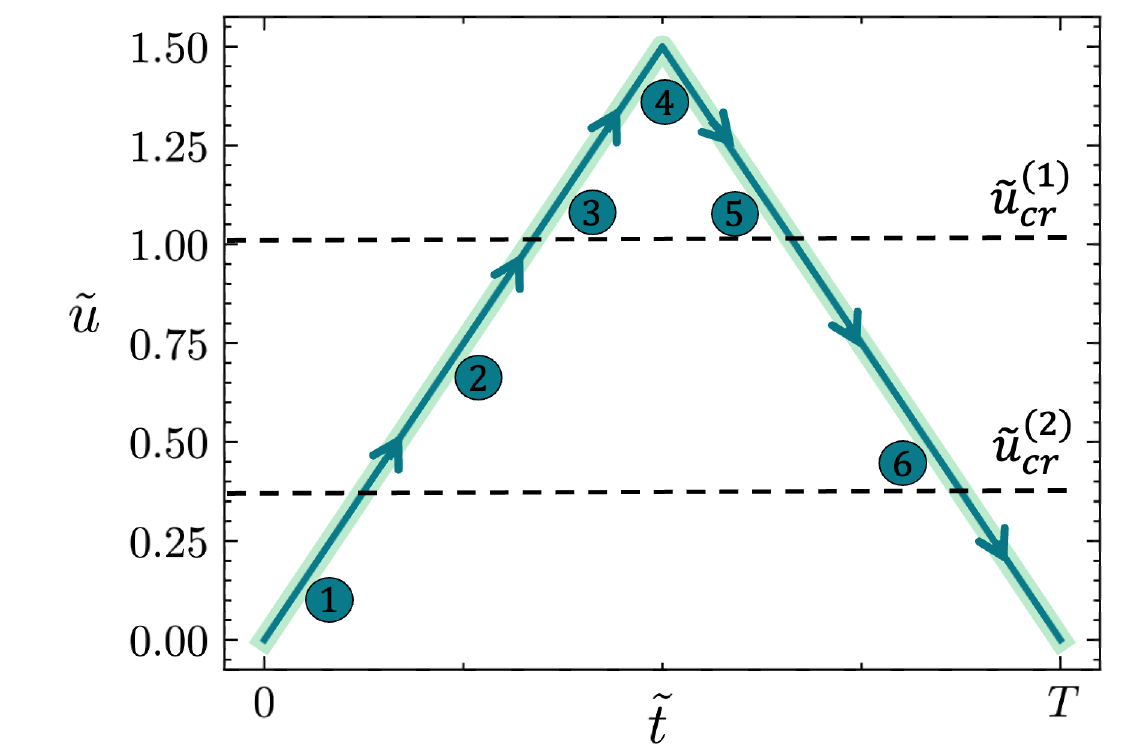}  \\ 
  (b)  \includegraphics[width=0.9\linewidth]{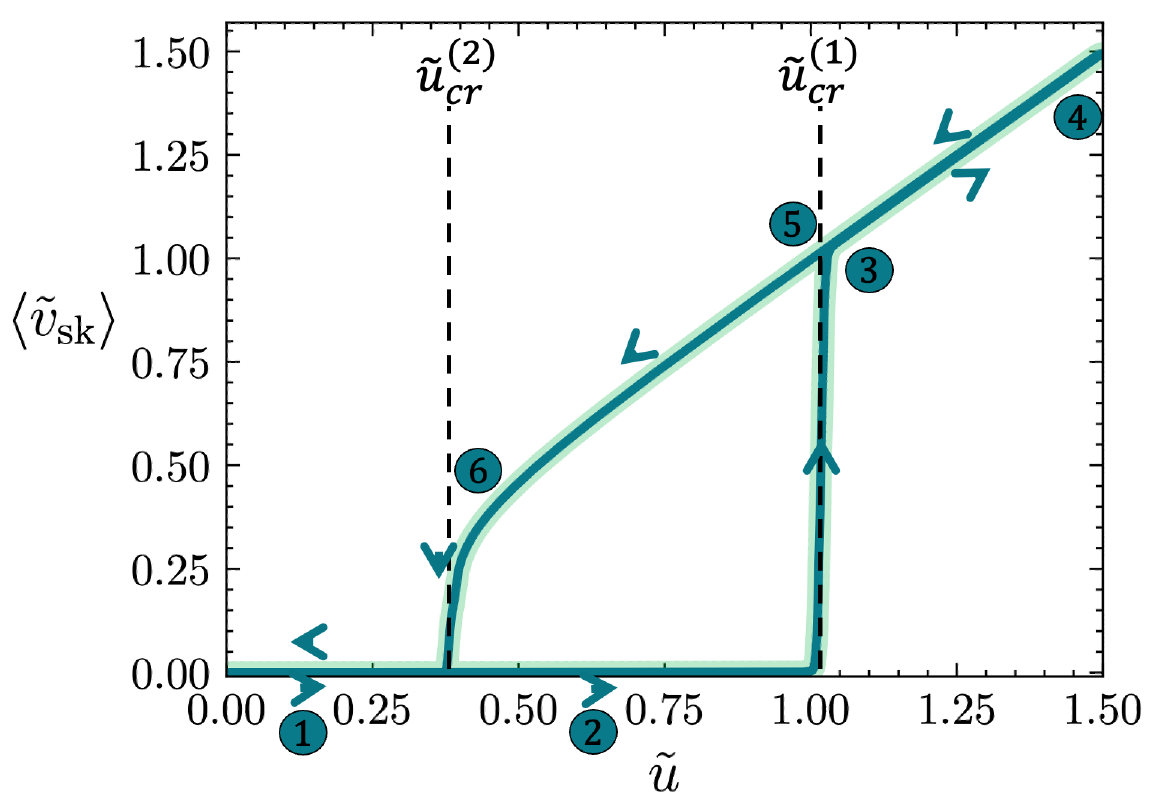}   
      \caption{(a) Protocol for changing the push force $\tilde u(\tilde t)$ leading to hysteresis: the push force is slowly ramped up and then gradually decreased back to zero over some large time $T$.  (b) The numerically calculated skyrmion velocity $\langle \tilde v_{\rm sk}\rangle$ exhibits hysteresis. The arrows and the numerical labels can be used as a guide to the eye to compare the two panels.}
      \label{fig:hysteresis}
\end{figure}
Another noteworthy feature of massive dynamics $\tilde m_\mathrm{sk} \gg 1$ is hysteresis of the $\tilde v_{\mathrm{sk}}$-vs-$\tilde u$ curve, which we consider below. For the discussion below, let us summarize the two critical push forces, found above, as 
\begin{equation}
\begin{aligned}
   & \tilde u_\mathrm{cr}^{(1)} = 1  \\
   & \tilde u_\mathrm{cr}^{(2)} = \frac 4{\pi\sqrt{\tilde m_\mathrm{sk}}},
\end{aligned} \label{critical_currents}
\end{equation}
where $\tilde u_\mathrm{cr}^{(1)} > \tilde u_\mathrm{cr}^{(2)}$ for $\tilde m_\mathrm{sk} \gg 1$. According to Fig.~\ref{fig:washerboard}, the push forces above the first critical current $\tilde u > \tilde u_{\mathrm{cr}}^{(1)}$ correspond to monotonically decreasing potential causing the skyrmion to propagate with a finite velocity. For lower push forces in the window between the two critical values $\tilde u_{\mathrm{cr}}^{(1)}>\tilde u > \tilde u_{\mathrm{cr}}^{(2)}$, the washboard potential contains local minima and admits two solutions. (i) One solution corresponds to a stationary skyrmion parked at one of the local minima, e.g. $\tilde x(\tilde t) = 0$. (ii) Another solution, which was discussed in Sec.~\ref{sec:large_lambda} above, corresponds to a propagating solution. Whether the skyrmion takes either of the two solutions (i) or (ii), depends on the history of $\tilde u(t)$, which leads to hysteresis. Specifically, we consider the following driving protocol for the push force
\begin{align}
    \tilde u (t) = 1.5 \left[1 - \left|\frac{2t}{T}-1\right|\right] \label{drive_protocol}
\end{align}
over the time interval $0<t<T$. The protocol  $\tilde u(t)$ is illustrated in Fig.~\ref{fig:hysteresis}(a). By taking sufficiently large $T$ compared to the fluctuation time scale by the disorder, the push force is slowly ramped up from $0$ to a maximum value of $\tilde u = 1.5 > \tilde u_\mathrm{cr}^{(1)}$ followed by a slow reduction back to $0$. We substitute Eq.~(\ref{drive_protocol}) into the Eq.~(\ref{1d_eom_cosine_disorder_dimensionless}), and solve it numerically and average skyrmion velocity $\langle \tilde v_\mathrm{sk} \rangle$ over a sufficiently long-time compared with the oscillation time of the skyrmion but small compared to $T$. The $\langle \tilde v_\mathrm{sk} \rangle$-vs-$\tilde u$ curve has hysteresis as shown in  Fig.~\ref{fig:hysteresis}(b). Initially, at $0 < t < T/2$, the stationary skyrmion remains pinned by the minima of the washboard potential at $\tilde u< \tilde u_{\mathrm{cr}}^{(1)}$. Eventually, $\tilde u$ surpasses $\tilde u_{\mathrm{cr}}^{(1)}$ and the minima disappear, leading to an abrupt jump to the propagating solution. For later times, at $T/2 < t < T$, the push force decreases, but the skyrmion sustains a non-zero velocity $\langle \tilde v_{\rm sk} \rangle$ even for low push forces at $\tilde u_{\mathrm{cr}}^{(2)}< \tilde u < \tilde u_{\mathrm{cr}}^{(1)}$ due to its inertia. Eventually, at $\tilde u < \tilde u_{\mathrm{cr}}^{(2)}$, the potential minima are strong enough to pin the skyrmion.


\section{Conclusion}
\label{sec:conclusion}

In this work, we show that skyrmions confined to quasi-1D geometries generically acquire a non-zero mass. We investigate the dynamics of a massive skyrmion pushed by electric current $\tilde u$ in 1D. We find that the dynamics can be classified by a dimensionless parameter $\tilde m_\mathrm{sk}$, given by Eq.~(\ref{lambda}). In the overdamped regime $\tilde m_\mathrm{sk} \ll 1$, the skyrmion velocity scales as a square root $\tilde v_{\rm sk}\propto \sqrt{\tilde u-\tilde u_\mathrm{cr}}$ close to depinning current $\tilde u_\mathrm{cr}$. In contrast, in the overdamped regime, the skyrmion velocity exhibits a sharp step-like onset at $\tilde u_\mathrm{cr}$ as shown in Fig.~\ref{fig:u_vs_v}. A similar step-like feature was recently observed in Ref.~\cite{Song2024}. Remarkably, in the underdamped regime $\tilde m_\mathrm{sk} \gg 1$, non-zero skyrmion mass could lead to hysteresis of the current-driven skyrmion dynamics, see Fig.~\ref{fig:hysteresis}. The experimental realization of hysteresis in the dynamics of skyrmion confined in one dimension could have applications in skyrmion-based memory architectures.
\section*{Acknowledgements}
\label{sec:acknowledgements}

This work was supported by the Office of Basic Energy Sciences, Division of Materials Sciences and Engineering, U.S. Department of Energy, under Award No. DE-SC0020221. K.Takahashi was supported by the Summer Undergraduate Research Fellowships at the Hamel Center for Undergraduate Research of the University of New Hampshire.

\bibliography{bibliography}	

\appendix

\section{Derivation of the 1D equation of motion}
\label{sec:derivation_1d_eom}

We aim to isolate $x(t)$  from the Thiele equation~(\ref{Thiele_basic}), which is given by
\begin{align}
\bm{G} \times \left( \bm{u}(t) - \dot{\bm{r}}(t) \right) + \mathcal{D} \left( \beta \, \bm{u}(t) - \alpha  \, \dot{\bm{r}}(t) \right) - \bm\nabla U(\bm{r}(t)) =\bm{0} \nonumber
\end{align}
For the vector components, the velocity of skyrmion
\begin{align}
    \bm{v}(t) = \dot{x}(t) \hat{\bm{x}} + \dot{y}(t) \hat{\bm{y}}
\end{align}
and the velocity of conduction electrons
\begin{align}
    \bm{u}(t) = u_x(t) \hat{\bm{x}} + u_y(t) \hat{\bm{y}}
\end{align}
Here, our model is
\begin{align}
U &= -\frac{1}{2}k\, y(t)^2 + U_{\mathrm{dis}}(x(t))\\
\bm{u}(t) &= u(t) \, \hat{\bm{x}} ~~(u_y(t) = 0)
\end{align}
Therefore, Eq.~(\ref{Thiele_basic}) can be written as a system of ODEs
\begin{align}
\mathcal{G} \dot{y} + \mathcal{D} \beta u - \mathcal{D} \alpha  \dot{x} - \partial_x U_{\mathrm{dis}} = 0 \label{ODE1}\\ 
\mathcal{G} u - \mathcal{G} \dot{x} - \mathcal{D} \alpha \dot{y} - ky = 0 \label{ODE2}
\end{align}
Differentiating both Eq.~(\ref{ODE1}) $\&$ Eq.~(\ref{ODE2}) with respect to $t$,
\begin{align}
 \ddot{y} + \frac{\mathcal{D} \beta}{\mathcal{G}} \dot{u} - \frac{\mathcal{D} \alpha}{\mathcal{G}}  \ddot{x} - \frac{\partial_x^2 U_{\mathrm{dis}}}{\mathcal{G}} \dot{x} &= 0 \label{ODE3}\\
 \dot{u} - \ddot{x} - \frac{\mathcal{D} \alpha}{\mathcal{G}} \ddot{y} - \frac{k}{\mathcal{G}} \dot{y} &= 0 \label{ODE4}
\end{align}
Eq.~(\ref{ODE1}) $\&$ Eq.~(\ref{ODE3}) can be written
\begin{align}
\dot{y} &=  \frac{\mathcal{D} \alpha}{\mathcal{G}}  \dot{x} - \frac{\mathcal{D} \beta}{\mathcal{G}} u + \frac{\partial_x U_{\mathrm{dis}}}{\mathcal{G}} \label{ODE5}\\
\ddot{y} &= \frac{\mathcal{D} \alpha}{\mathcal{G}} \ddot{x} - \frac{\mathcal{D} \beta}{\mathcal{G}} \dot{u} + \frac{\partial_x^2 U_{\mathrm{dis}}}{\mathcal{G}} \dot{x} \label{ODE6}
\end{align}
Plugging in Eq.~(\ref{ODE5}) $\&$ Eq.~(\ref{ODE6}) into (\ref{ODE1}), one can get rid of $y(t)$ and get a differential equation of $x(t)$,
\begin{align}
\begin{split}
 {\mathcal{G}} \dot{u} - {\mathcal{G}} \ddot{x} &- \mathcal{D} \alpha \left( \frac{\mathcal{D} \alpha}{\mathcal{G}} \ddot{x} - \frac{\mathcal{D} \beta}{\mathcal{G}} \dot{u} + \frac{\partial_x^2 U_{\mathrm{dis}}}{\mathcal{G}} \dot{x} \right) \\
 &- k \left( \frac{\mathcal{D} \alpha}{\mathcal{G}}  \dot{x} - \frac{\mathcal{D} \beta}{\mathcal{G}} u + \frac{\partial_x U_{\mathrm{dis}}}{\mathcal{G}} \right)  = 0
\end{split}
\end{align}
Assuming $k$ is nonzero, we derive Eq.~(\ref{1d_eom}).
\\

\section{Derivation of current-velocity relation for strictly constricted 1D motion}
\label{sec:k_inf_limit}
Eq.~(\ref{1d_eom_massless}) is given by
\begin{align}
\frac{d\tilde x}{d\tilde t} =  \tilde u - \sin\tilde x. \nonumber
\end{align}
We separate $\tilde x$. We integrate over the one time period of motion $\tilde T$ with respect to $\tilde t$, and integration range with respect to $\tilde x$ is over the period $2 \pi$, given by
\begin{align}
\tilde T(\tilde u) &= \int_{0}^{2 \pi} d \tilde x~\frac{1}{ \tilde u - \sin(\tilde x)} \label{one_period}
\end{align}
This can be transformed into a complex integration along a path $C: z = e^{i \tilde x}~(0 \leq \tilde x \leq 2\pi)$,
\begin{align}
\tilde T(\tilde u) =& \int_{0}^{2 \pi} d \tilde x~\frac{1}{\tilde u - \sin(\tilde x)} \\
=& \oint_C \frac{dz}{iz}~\frac{1}{\tilde u - \frac{1}{2i} \left(z - \frac{1}{z}\right)} \nonumber \\
=& \oint_C dz~\frac{-2}{(z-i(\sqrt{\tilde u^2-1}+\tilde u))(z-i(-\sqrt{\tilde u^2-1}+\tilde u))}
\end{align}
Therefore, following the residue theorem, we get
\begin{align}
\tilde T(\tilde u) = \left\{
\begin{array}{ll}
\infty &  \text{if } 0 < \tilde u \leq 1\\[1em]
\frac{2 \pi}{\sqrt{\tilde u^2 - 1}} &  \text{if } 1 \leq \tilde u
\end{array}
\right. \label{integrate_b}
\end{align}
Immediately, we can get the average velocity of a skyrmion $\langle \tilde{v}_{\rm sk} \rangle$ in terms of $\tilde u$, given by
\begin{align}
\langle \tilde{v}_{\rm sk} \rangle &= \frac{2 \pi}{\tilde{T}(\tilde u)}
\end{align}
which leads to the relation~(\ref{current_velocity_large_friction}).

In the following, we obtain the asymptotic form of the large current density limit and the critical current density limit in (\ref{current_velocity_large_friction}). In the large current density limit where $\tilde u \gg 1$,
\begin{align}
\langle \tilde{v}_{\rm sk} \rangle \approx \tilde u
\end{align}
so that in the dimensional form,
\begin{align}
\langle v_{\rm sk} \rangle \approx \frac{\beta}{\alpha} \,u
\end{align}
In the vicinity of the critical current density $\tilde u_{\mathrm{cr}} = 1$, let $\tilde u = \tilde u_{\mathrm{cr}} + \tilde \epsilon$ where $ \tilde \epsilon \ll 1$,
\begin{align}
\langle \tilde v_{\rm sk} \rangle  &= \sqrt{\tilde u^2 - 1} = \sqrt{\left(\tilde u_{\mathrm{cr}} + \tilde \epsilon\right)^2 - 1} \nonumber \\ 
&\approx \sqrt{2 \tilde u_{\mathrm{cr}} \tilde \epsilon} = \sqrt{2 \tilde u_{\mathrm{cr}}} \left(\tilde u - \tilde u_{\mathrm{cr}} \right)^{1/2}
\end{align}

\end{document}